\documentclass[aps,prb,twocolumn,showpacs]{revtex4}
\usepackage{graphicx}
\usepackage{bm}

\begin{document}

\title{
Vanishing N\'eel Ordering of SU($n$) Heisenberg Model in Three Dimensions
}

\author{Noboru Fukushima}
\email{fukushima@lusi.uni-sb.de}
\affiliation{
Department of Physics, University of the Saarland,
66041 Saarbr\"ucken, Germany}

\begin{abstract}
The SU($n$) Heisenberg model represented by exchange operators is studied by
means of high-temperature series expansion in three dimensions, where
$n$ is an arbitrary positive integer.
The spin-spin correlation function and its Fourier transform $S(\mathbf{q})$
is derived up to $O[(\beta J)^{10}]$ with $\beta J$ being the
nearest-neighbor antiferromagnetic exchange in units of temperature.
The temperature dependence of $S(\mathbf{q})$ and next-nearest-neighbor
spin-spin correlation in the large $n$ cases show that dominant correlation
deviates from $\mathbf{q}=(\pi,\pi,\pi)$ at low temperature, which is
qualitatively similar to that of this model in one dimension.
The N\'eel temperature of SU(2) case is precisely estimated by analyzing
the divergence of $S(\pi,\pi,\pi)$.  Then, we generalize $n$ of SU($n$) to a
continuous variable and gradually increases from $n=2$.  We conclude
that the N\'eel ordering disappears for $n>2$.
\end{abstract}


\maketitle

\section{Introduction}

It is known that properties of quantum spin systems tend to approach
those of their corresponding classical spin systems as the spin magnitude
increases.
However, this is not necessarily the case for a sequence of models in
which the number of multipolar-interaction terms increases as the spin
magnitude increases.  Consequently the higher-spin systems may have
stronger quantum effects in this case.
In other words, such additional terms can break a classical
correspondence down even in high dimensions.
These ``large-spin-magnitude'' systems mentioned above include systems
in which one unit has more than two degrees of freedom, such as
orbitally degenerate systems\cite{Kugel82, FK,KF}.  In such
systems, many coupling constants appear in general.  However,
experimental information about multipolar couplings is limited.  As a
starting point to explore such systems, understanding of one of the
extreme limits must be useful.  Therefore in this paper, we investigate
an SU($n$) symmetric case
\cite{Uimin70,Sutherland75,Klumper99,Affleck86,Kawakami92,Batchelor03,
Yamashita98,Frischmuth99,Fukushima02,Fukushima03,
lima,vdb,RVBsu4,fermiMF,shen,Penc03,
Ohkawa85,Shiina97,chen72}
in three dimensions, where $n$ is the total
number of internal degrees of freedom, by means of high-temperature
series expansion (HTSE).

We consider a simple cubic lattice. Let each site take one of the $n$
colors denoted by $|\alpha\rangle$ with $\alpha = 1, 2, \cdots, n$.  Using
the Hubbard operator $X^{\alpha\beta} := |\alpha\rangle\langle\beta|$,
the exchange operator is expressed as
\begin{equation}
 P_{\mathbf{i},\mathbf{j}} := \sum_{\alpha=1}^n  \sum_{\beta=1}^n
X_\mathbf{i}^{\alpha\beta} X_\mathbf{j}^{\beta\alpha}.
\end{equation}
Colors of sites $\mathbf{i}$ and $\mathbf{j}$ are exchanged when
$P_{\mathbf{i},\mathbf{j}}$ is applied.
Then, an SU($n$) symmetric Hamiltonian reads
\begin{equation}
 {\cal H} :=  J \sum_{\langle \mathbf{i},\mathbf{j}\rangle} P_{\mathbf{i},\mathbf{j}},
\label{hamil}
\end{equation}
where the summation is taken over all the nearest neighbor pairs.
We consider the antiferromagnetic case, $J>0$.
Let us show the relations with spin operators explicitly for
some of the special cases below.

(i) When $n=2$, this model is reduced to the ordinary Heisenberg model
with $s=1/2$ by relation
$2P_{\mathbf{i},\mathbf{j}}-1=4 \bm{s}_\mathbf{i}\cdot \bm{s}_\mathbf{j}$.

(ii) The SU(3) case corresponds to $s=1$,
competing quadratic and biquadratic exchange interaction,\cite{PaPanicolaou88}
\begin{equation}
1+P_{\mathbf{i},\mathbf{j}} =
\bm{s}_\mathbf{i}\cdot
\bm{s}_\mathbf{j}+(\bm{s}_\mathbf{i}\cdot \bm{s}_\mathbf{j})^2.
\label{eq:su3rep}
\end{equation}

(iii) The SU(4) Heisenberg model is
related to spin 3/2 systems but more often discussed in the context of
orbital- and spin-degenerate systems\cite{Yamashita98,Frischmuth99,
Fukushima02,Fukushima03,
lima,vdb,RVBsu4,fermiMF,shen,Penc03,
Ohkawa85,Shiina97}.
In particular, the model in three dimensions has been used
as an effective model of CeB$_6$ to explain magnetic-field dependence of
the transition temperature of an antiferro-orbital ordering.
\cite{Ohkawa85,Shiina97}
The four local states
can be represented by
$|+)|+]$, $|+)|-]$, $|-)|+]$, $|-)|-]$,
where $|\pm)$ and $|\pm]$ represent an orbital state and a spin
state, respectively. The pseudo-spin operators are defined
by $t^z|\pm)=\pm\frac12|\pm)$, $t^\pm|\mp)=|\pm)$,
$s^z|\pm]=\pm\frac12|\pm]$, $s^\pm|\mp]=|\pm]$, and
the exchange operator for $n=4$ is rewritten as
\begin{equation}
P_{\mathbf{i},\mathbf{j}} = \bm{t}_\mathbf{i}
\cdot\bm{t}_\mathbf{j} + \bm{s}_\mathbf{i}\cdot \bm{s}_\mathbf{j}
  +4 (\bm{t}_\mathbf{i}\cdot\bm{t}_\mathbf{j})  (\bm{s}_\mathbf{i}\cdot \bm{s}_\mathbf{j})
 +\frac{1}{4}.
\end{equation}
The Pauli matrices $\bm{\tau}=2\bm{t}$ and $\bm{\sigma} =2\bm{s}$
may simplify this expression, {\em i.e.},
\begin{equation}
4 P_{\mathbf{i},\mathbf{j}}
= \bm{\tau}_\mathbf{i}\cdot\bm{\tau}_\mathbf{j}
+ \bm{\sigma}_\mathbf{i}\cdot \bm{\sigma}_\mathbf{j}
  + (\bm{\tau}_\mathbf{i}\cdot\bm{\tau}_\mathbf{j})  (\bm{\sigma}_\mathbf{i}\cdot \bm{\sigma}_\mathbf{j})
 +1.
\label{paulirep}
\end{equation}

Note that there is a different representation of SU($n$) Heisenberg model
studied in detail using Quantum Monte Carlo method
(QMC)\cite{Harada02}.  However, the QMC for the Hamiltonian
(\ref{hamil}) suffers from minus sign problems in more than one
dimensions \cite{Harada02, Frischmuth99}

If the number of competing order parameters is large and frustration
exists, the transition temperature can greatly be reduced from the
mean-field value even in three dimensions. That will be the case with the
Hamiltonian (\ref{hamil}).
First, it is isotropic with respect to $n^2-1$ independent interacting
components.  Furthermore, it contains frustration as most clearly seen
in Eq.~(\ref{paulirep}) of the SU(4) case.  Each of the 15 components,
$\tau^\alpha$, $\sigma^\beta$, $\tau^\gamma \sigma^\delta$, attempts an
antiparallel alignment.  However, it cannot be attained simultaneously,
{\em e.g.}, simultaneous N\'{e}el states of $\tau^z$ and $\sigma^z$ produce a
ferromagnetic alignment of the product $\tau^z \sigma^z$.  Such
frustration is not so explicit in the SU(3) case in Eq.~(\ref{eq:su3rep})
as in the SU(4) case, yet appearance of the minus-sign problem in the
QMC indicates that similar frustration lies in it.\cite{Harada02}

This frustration should become stronger as $n$ increases.  One can see
in one dimension how the system finds a compromise against this
frustration.  The SU($n$) Heisenberg model in one dimension can be
exactly solved.\cite{Uimin70,Sutherland75,Klumper99, Affleck86,
Kawakami92,Batchelor03} An important point is that the ground state has a
quasi--$n$-site periodicity.  The SU(4) case is numerically studied in
more detail; ground-state properties by the density matrix
renormalization group\cite{Yamashita98}, and thermodynamic properties by
the QMC\cite{Frischmuth99} and the HTSE\cite{Fukushima02,Fukushima03}.
The real-space spin-spin correlation function as a function of spin-spin
distance has a positive value every four sites.  Its Fourier transform
has a outstanding cusp at $q=\pi/2$ and a small cusp at $q=\pi$.
Regarding the temperature dependence, as temperature decreases
correlation with two-site periodicity develops, and then at lower
temperature another correlation occurs with $n$-site periodicity.
\cite{Fukushima02}

Such peculiar correlation in one dimension could suggest an exotic
ordering in higher dimensions. However, not much is known about the
antiferromagnetic SU($n$) Heisenberg model in three dimensions.
In fact, its ferromagnetic variant is studied in Ref.~\onlinecite{chen72}
by the HTSE for the uniform susceptibility that needs less effort to be
calculated than susceptibility of other wave-numbers.
To our knowledge, our calculation in this paper is the first HTSE aiming at
antiferromagnetic exchange of this model.
In this study, we investigate the model systematically by changing
parameter $n$ of SU($n$), and report some unique features of spatial
correlation at finite temperature.  Namely, in Sec.~\ref{sec:corfunc} we
show that the behavior of correlation functions is similar to that in one
dimension, and in Sec.~\ref{sec:neeltem} that the N\'eel temperature
disappears as $n$ increases from $n=2$.

\section{Temperature dependence of correlation functions}
\label{sec:corfunc}

What should be calculated here is $\langle P_{\mathbf{i},\mathbf{j}} \rangle$,
which is related to a correlation function.
For example, when $n=4$, fifteen components in Eq.~(\ref{paulirep})
contribute equally, and thus
$
\langle \tau_\mathbf{i}^\alpha \tau_\mathbf{j}^\alpha\rangle
=
\langle \sigma_\mathbf{i}^\beta \sigma_\mathbf{j}^\beta\rangle
=\langle \tau_\mathbf{i}^\gamma \sigma_\mathbf{i}^\delta
         \tau_\mathbf{j}^\gamma \sigma_\mathbf{j}^\delta \rangle
=
\langle 4 P_{\mathbf{i},\mathbf{j}} -1 \rangle /15 $.
For general $n$, we can define correlation function between sites
$\mathbf{i}$ and $\mathbf{j}$ as
\begin{equation}
S_{\mathbf{i}-\mathbf{j}} := \langle X_\mathbf{i}^{\alpha\beta} X_\mathbf{j}^{\beta\alpha} \rangle=
\frac{1}{n^2-1} \left( \langle  P_{\mathbf{i},\mathbf{j}} \rangle -\frac{1}{n} \right),
\end{equation}
for $\mathbf{i}\neq \mathbf{j}, \alpha \neq \beta$, which does not
depend on $\alpha$ nor $\beta$ because of the SU($n$) symmetry.
Its Fourier transform is denoted by $S(\mathbf{q})$.

The high-temperature expansion
is performed by expanding
the Boltzmann factor $e^{-\beta {\cal H}}$
in $\beta$.
In practice, the series coefficients in the thermodynamic limit
are exactly obtained by a linked-cluster expansion.\cite{Domb3}
To obtain the series expansion of
$\langle P_{i,j} \rangle$ up to $O[(\beta J)^M]$,
we need to calculate ${\rm Tr}[({\cal H}_{\rm L})^m]$ and
${\rm Tr}[P_{i,j}({\cal H}_{\rm L})^m]$ for $0\le m\le M$,
where ${\cal H}_{\rm L}$ is the Hamiltonian in a linked cluster.
In calculating the traces, we use
a property of the permutation operator, and
calculation of the traces is reduced to a combinatorial problem
to count the number of circular permutations in a product of permutations.
\cite{Handscomb64,chen72,Fukushima03}
Here, an important point of our analysis is that the series coefficients are
obtained as polynomials of $n$, for example,
\begin{eqnarray}
S(\pi,\pi,\pi)&=&
\frac{1}{{n}} + \frac{6}{{ n^2}}(\beta J) + \frac{36}{{ n^3}}(\beta J)^2
\nonumber \\
&&+ \frac{\left( 216 - 22\,{ n^2} \right)}{{ n^4}}(\beta J)^3 +\ldots.
\end{eqnarray}
Namely, the order of the series for every $n$ is the same.
We have
obtained the series for $S_{\mathbf{i}-\mathbf{j}}$ up to $O[(\beta
J)^{9}]$ for arbitrary $\mathbf{i}- \mathbf{j}$, and consequently
$S(\mathbf{q})$ up to $O[(\beta J)^{9}]$ for arbitrary $\mathbf{q}$.  As a
special interest, $S(\pi,\pi,\pi)$ is obtained up to $O[(\beta
J)^{10}]$. The series are extrapolated using the Pad\'{e} approximation (PA).

\begin{figure}\begin{center}
\includegraphics[width= 8cm]{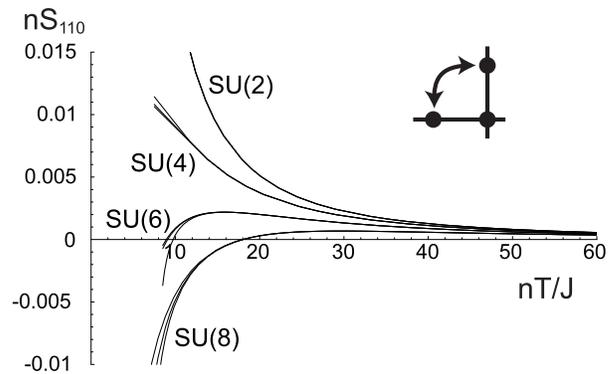}
\end{center}
\caption{\label{fig:nncor}
Temperature dependence of the next-nearest-neighbor correlation function,
$S_{\mathbf{i}-\mathbf{j}}$ with $\mathbf{i}-\mathbf{j}=(110)$.
}
\end{figure}

First of all, in order to see the temperature-dependent nature of the
spatial correlation, we analyze the \textit{next}-nearest-neighbor
correlation function $S_{110}$ as we have done in one dimension in
Ref.~\onlinecite{Fukushima02}.  Figure \ref{fig:nncor} shows the
results.  Both the axes are scaled so that the high-temperature limit of
every $n$ matches.
Here, we have simultaneously plotted extrapolation from several
different choices of the PA, and the difference between them
approximately represents an error of the extrapolation.

The lowest order of the series of $S_{\mathbf{i}-\mathbf{j}}$ has
a N\'eel-order-type correlation in any $n$.
That is, the series of $S_{i_xi_yi_z}$ starts with
$O[(\beta J)^{|i_x|+|i_y|+|i_z|}]$ with sign $(-1)^{|i_x|+|i_y|+|i_z|}$.
Therefore, $S_{110}>0$ at high temperature.
However, as antiferromagnetic correlation of each interacting component
becomes larger, each short-range order disturbs another because of the
frustration.  The change of the sign of $S_{110}$ suggests that the
correlation acquire a longer period at low temperature.  For $n\ge6$, it
is clear that $S_{110}$ changes the sign at a low temperature that
increases with $n$ in this scale.  For smaller $n$, the relevant
temperature range is below the converged region, and it is difficult to
conclude from this data.

Such a change of correlation at low temperature also appears in the
Fourier transform of the correlation function.
A naive extrapolation of the series for $S(\mathbf{q})$ shows bad
convergence for several $\mathbf{q}$.
This is probably because information at position $\mathbf{x}$
is lacking when $\cos(\mathbf{x}\cdot \mathbf{q})\simeq 0$.
In order to avoid it, we extrapolate the series of a complex function
$ \sum_{\mathbf{x}\ge0}
\langle X_j^{\alpha\beta} X_{j+x}^{\beta\alpha} \rangle
e^{-i\mathbf{x}\cdot\mathbf{q}}  w(\mathbf{x})$,
and after that take the real part of the extrapolated function.
Here, the summation is taken for $x\ge0$, $y\ge0$, $z\ge0$, and
$w(\mathbf{x})$ is two to the power of the number of nonzeros in
$x,y,z$.
This method also makes use of its imaginary part, and the convergence
becomes better than the naive extrapolation.

\begin{figure}
\begin{center}\includegraphics[  width=8cm
]{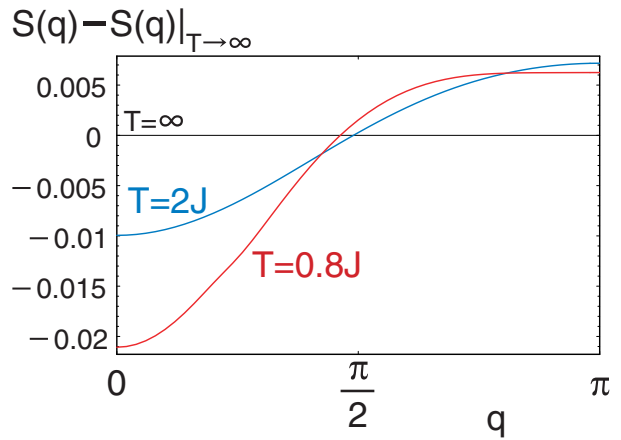}\end{center}
\caption{\label{fig:sq}
The Fourier transform of the correlation function
as a difference from its high-temperature limit,
$ S(q,q,q) - S(\mathbf{q})|_{T\rightarrow\infty}$,
for the SU(16) model.
}
\end{figure}

\begin{figure}
\begin{center}
\includegraphics[width= 8cm]{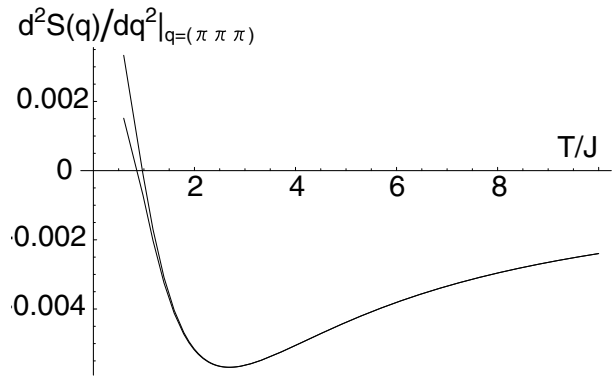}
\end{center}
\caption{\label{fig:ddsqn16}
Temperature dependence of
${{\rm d}^2 S(q,q,q)}/{{\rm d}q^2}$ at $q=\pi$
for the SU(16) model by two different PAs.
}
\end{figure}

As we have seen in the
next-nearest-neighbor correlation function, the analysis becomes easier
as $n$ increases.  Therefore, we show $S(\mathbf{q})$ of the SU(16) case
in Fig.~\ref{fig:sq} for a diagonal direction in the $\mathbf{q}$-space,
$ S(q,q,q) $.  Here, its high temperature limit, which does not
depend on $\mathbf{q}$, is subtracted.
As temperature decreases, antiferromagnetic correlation develops and
$S(\pi,\pi,\pi)$ increases as the curve at $T=2J$ shows.  However, with
further decrease of temperature, $S(\pi,\pi,\pi)$ starts decreasing and
$S(\mathbf{q})$ with other $\mathbf{q}$ increases.  It is not very clear
from Fig.~\ref{fig:sq} if the maximum of $S(\mathbf{q})$ starts moving
from $(\pi,\pi,\pi)$. Hence, let us show another quantity.  Note that if
the second derivative of $S(\mathbf{q})$ at $\mathbf{q}=(\pi,\pi,\pi)$
changes sign, the position of the maximum clearly moves.
Its temperature dependence for the SU(16) model is plotted in
Fig.~\ref{fig:ddsqn16}.
It shows a change of the sign around $T\sim J$. In fact, this
temperature of changing sign seems to hardly depend on $n$ of SU($n$).
The result above suggests that the N\'eel order disappear at least in
the SU($n$) model with large $n$. Namely, order with wave number
$\mathbf{q}\neq(\pi,\pi,\pi)$, or disorder, should appear.
Then, the next question is, at which $n$ the N\'eel order disappears.

\section{N\'eel Temperature}
\label{sec:neeltem}

It is known that the SU(2) Heisenberg model has the N\'eel order
at low temperature. Therefore in this section, we gradually
increases $n$ of SU($n$) from $n=2$.
As a preparation for that,
we find a reliable way to estimate a transition temperature first,
using the SU(2) model.
The N\'eel temperature $T_{\rm N}$ can be characterized by divergence of
$S(\pi,\pi,\pi)$, namely, by a singularity of $S(\pi,\pi,\pi)$ as a
function of $\beta J$.  In order to analyze such a singularity, we use
so-called D-log-Pad\'e approximation (DLPA), {\em i.e.}, the PA for the
logarithmic derivative.
In using the DLPA, transformation of the
expansion variable may improve the convergence of extrapolation.
We choose a transformation\cite{Domb13,Pan99}
\begin{equation}
\beta J=\frac{x}{1-{ a}^2 x^2},
\label{eq:trans}
\end{equation}
where $a$ is an adjustable parameter that improves convergence.  In
fact, the singularity closest to the origin in the original series is
near the imaginary axis, while $T_{\rm N}$ corresponds to a
singularity on the real axis.  With this transformation, singularities
on the real axis approach to the origin and those near the imaginary
axis go away.  Since the DLPA can estimate a position of the nearest
singularity most accurately, errors of the DLPA can become smaller by
this transformation.

\begin{figure}
\begin{center}\includegraphics[  width=8cm
]{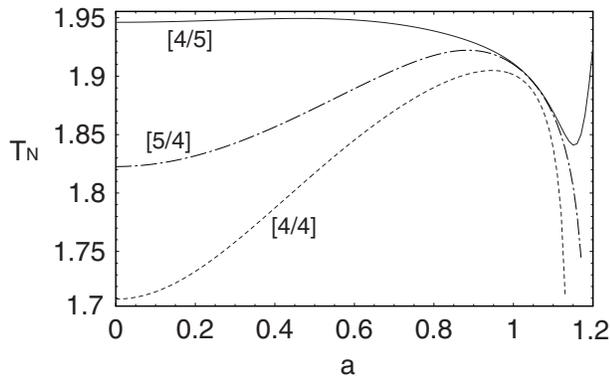}
\caption{\label{fig:valtran}
The N\'eel Temperature $T_{\rm N}$ of the SU(2) Heisenberg model
obtained by three different DLPAs as a function of the parameter $a$.
}
\end{center}\end{figure}

In order to find an optimal $a$, we calculate $T_{\rm N}$ as a function
 of $a$ for a couple of choices of the DLPA, and we adopt $a$ at which
 difference among different DLPAs is the smallest.  Figure
 \ref{fig:valtran} shows $T_{\rm N}$ as a function of $a$ for three
 different choices of the DLPA.  Here, $[m/n]$ denotes the DLPA with a
 polynomial of order $m$ over a polynomial of order $n$.  Since [4/4] is
 from the series one-order lower than the others, we choose $a$ at which
 [5/4] and [4/5] are the closest, namely, $a=1.04$.  Then, $T_{\rm N}/J$
 obtained here is 1.898, which is close to those in the literature,
 $1.892$ by the QMC\cite{Sandvik98} and $1.888$ by the HTSE\cite{Oitmaa}.
In addition, we have obtained a critical exponent simultaneously.
We assume that the critical exponent does not depend on the spin magnitude,
and compare it with that of the {\em classical} antiferromagnetic Heisenberg model,
in which $S(\pi,\pi,\pi)$ is identical to the staggered susceptibility
equivalent to the `uniform susceptibility of the ferromagnetic model'.
Hence we can compare the critical exponent $\gamma$.
The estimation from our calculation above is $\gamma=1.399$,
which is close to
1.396 by a Monte Carlo method\cite{Campostrini02},
1.406 by HTSE\cite{Butera97}, and 1.388 by a field theory\cite{Jasch01},
of the classical O(3) model.
Therefore, we trust this way of analyzing $T_{\rm N}$,
and use it also for the SU($n$) model with $n>2$.
Also for $n>2$,
the variable transformation (\ref{eq:trans}) should work
because the singularity closest to the origin in the original series is
near the imaginary axis.

\begin{figure}
\begin{center}\includegraphics[  width=8cm
]{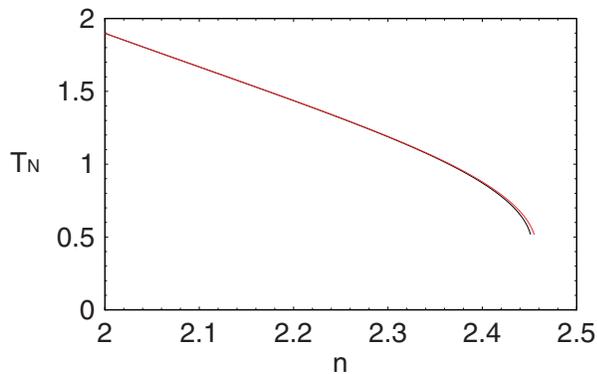}
\caption{\label{fig:ndep}
The N\'eel Temperature $T_{\rm N}$ of the SU($n$) Heisenberg model
as a function of $n$ with $a=1.04$.
Here, [5/4] and [4/5] are plotted together.
}
\end{center}\end{figure}

Originally $n$ is an integer because it is the number of internal
degrees of freedom.  However, after obtaining the series as an analytic
function of $n$, we can regard $n$ as a continuous variable that has a
physical meaning only when it happens to take integer values.  In fact,
as $n$ changes continuously, the properties of the series change also
continuously. Therefore, we gradually increases $n$ from $n=2$ and see
$n$-dependence of $T_{\rm N}$.  Figure \ref{fig:ndep} shows the results.
Here, we fix $a=1.04$, and we plot [5/4] and [4/5] together.  The
optimal $a$ may depend on $n$.  However, since the difference between
the two curves is very small, we regard it as an optimal $a$.  As $n$
increases, $T_{\rm N}$ decreases almost linearly, and at $n\sim2.45$ the
singularity corresponding to $T_{\rm N}$ runs away from the real axis to
a complex value with a finite imaginary part. For larger $n$, we do
not find any singularity on the antiferromagnetic side of the real axis.
Therefore, we conclude that the value of $n$ at which
N\'eel order disappears lies in the range $2<n<3$.

In addition, we have also analyzed $S(\mathbf{q})$ with
$\mathbf{q}\neq(\pi,\pi,\pi)$. However, we have not found any
symptom of ordering with $\mathbf{q}\neq(\pi,\pi,\pi)$ in the
temperature region that we can reach by the present order of the series.

\section{summary}

In summary, we have performed high-temperature series expansions for
the SU($n$) Heisenberg model in three dimensions with arbitrary $n$.
First of all,
we have calculated a \textit{next}-nearest-neighbor correlation
function.  At least at large $n$, it changes the sign at low
temperature, which suggests that the correlation should not be like N\'eel
order, but have a longer period. Analysis of the Fourier transform of the
correlation function also supports that the ground state of the large-$n$
SU($n$) Heisenberg model does not have the N\'eel-order correlation.
Then, we have turned to an approach from $n=2$.  Since the SU(2)
Heisenberg model has the N\'eel order, it should disappear at a
certain $n$.  We have first found a reliable way to estimate a
transition temperature by analyzing the divergence of the $(\pi,\pi,\pi)$
component of the correlation function.
Next, we generalize $n$ to a continuous
variable and increase $n$ gradually from $n=2$.
We have concluded that the N\'eel
ordering disappears for $n>2$.

\acknowledgments
The author would like to thank Y.~Kuramoto for
directing him to this topic and for stimulating discussions.
He also appreciates useful comments on the manuscript from A.~Honecker.
This work was supported in part by the Deutsche Forschungsgemeinschaft (DFG).
Early stages of this work were supported by the Japan Society for the
Promotion of Science and by the Max-Planck Institute for the Physics of
Complex Systems.
Different stages of this work were supported by the DFG
and the Technical University Braunschweig.
Parts of the numerical calculations were performed on
the {\tt cfgauss} at the computing centers of the TU Braunschweig.

\end{document}